# Dynamic management of transactions in distributed real-time processing system

Y. Jayanta Singh[1] , Yumnam Somananda Singh[2],  Ashok Gaikwad[2] and S.C. Mehrotra[3]

[1]Faculty of Information Technology,
7[th] October University, Misurata, Libya
y_jayanta@yahoo.com

[2]Faculty of Computer Application
Institute of Management Studies & I. T (IMSIT), India
ysomananda@gmail.com

[3]Dept of Information Technology and Computer Science
Dr. B. A. Marathwada University, Aurangabad, India
mehrotrasc@rediffmail.com

*ABSTRACT*

*Managing the transactions in real time distributed computing system is not easy, as it has heterogeneously networked computers to solve a single problem. If a transaction runs across some different sites, it may commit at some sites and may failure at another site, leading to an inconsistent transaction. The complexity is increase in real time applications by placing deadlines on the response time of the database system and transactions processing. Such a system needs to process Transactions before these deadlines expired. A series of simulation study have been performed to analyze the performance under different transaction management under conditions such as different workloads, distribution methods, execution mode-distribution and parallel etc. The scheduling of data accesses are done in order to meet their deadlines and to minimize the number of transactions that missed deadlines. A new concept is introduced to manage the transactions in dynamic ways rather than setting computing parameters in static ways. With this approach, the system gives a significant improvement in performance.*

*KEYWORDS*

*Real time system, dynamic management, intelligent agents*

## 1. INTRODUCTION

Real-Time Database Systems are becoming increasingly important in a wide range of applications, such as telecommunications, mobile communication systems, nuclear reactor control, traffic control systems, computer integrated manufacturing, robotics and military systems. A Real-Time Database System (RTDBS) is a transaction processing system that is designed to handle workloads where transactions have deadlines. The objective of the system is to meet these deadlines. As the world gets smarter and more informatics, demands on IT will grow. Many converging technologies are coming up like emerging IT delivery model-cloud computing. Demands of the real time distributed database are also increasing. Many transaction complexities are there in handling concurrency control and database recovery in distributed database systems. Two-phase commit protocol is most widely used to solve these problems [1]. To ensure such transaction atomicity, commit protocols are implemented in distributed database system. A uniform commitment is guarantee by a commit protocol in a distributed transaction execution to ensure that all the participating sites agree on a final outcome. Result may be either a commit or an abort condition.

Many real time database applications in areas of communication system and military systems are distributed in nature. In a real time database system the transaction processing system that is designed to handle workloads where transactions have complete deadlines. To ensure transaction atomicity, commit protocol are implemented in distributed database system. Experimental performances of transaction scheduling under variety of workloads and different system configuration are evaluated through this simulation. Many database researchers have proposed varieties of commit protocols like Two phase commit and Nested two phase commit [2, 3], Presumed commit [4] and Presume abort [3], Broadcast Two phase commit , Three phase commit [5,6] etc. These require exchanges of multiple messages, in multiple phases, between the participating sites where the distributed transaction executed. Several log records are generated to make permanent changed to the data disk, demanding some more transaction execution time [4, 7, 8]. Proper scheduling of transactions and management of its execution time are important factors in designing such systems.

Transactions processing in any database systems can have real time constraints. The scheduling transactions with deadlines on a single processor memory resident database system have been developed and evaluated the scheduling through simulation [9]. A real time database system is a Transaction processing system that designed to handle workloads where transactions have complete deadlines. In case of faults, it is not possible to provide such guarantee. Real actions such as firing a weapon or dispensing cash may not be compensatable at all [10]. Proper scheduling of transactions and management of its execution time are the important factors in designing such systems. In such a database, the performance of the commit protocol is usually measured in terms of number of transactions that complete before their deadlines. The transaction that miss their deadlines before the completion of processing are just killed or aborted and discarded from the system without being executed to completion [11].

The section 2 describes the simulation details. In section 3, simulation model and simulation parameters are given. The detail experiments, results and analysis are given in section 4. The overall conclusions are discussed in section 5.

## 2. SIMULATION DETAILS

This study is in continuation of our work in the same domain [12,13]. The study follows the real time processing model [14,15,16] and transaction processing addressing timeliness [17]. This model has six components: (i) a source (ii) a transaction manager (iii) a concurrency control manager (iv) a resource manager (v) a recovery manager (vi) a sink to collects statistics on the completed transactions. A network manager models the behavior of the communications network. The definitions of the components of the model are given below.

(i) The source:
This component is responsible for generating the workloads for a site. The workloads are characterized in terms of files that they access and number of pages that they access and also update of a file.

(ii) The transaction manager:
The transaction manager is responsible for accepting transaction from the source and modeling their execution. This deals with the execution behavior of the transaction. Each transaction in the workload has a general structure consist of a master process and a number of cohorts. The master resides at the sites where the transaction was submitted. Each cohort makes a sequence of read and writes requests to files that are stored at its sites. A transaction has one cohort at each site where it needs to access data.

To choose the execution sites for a transaction's cohorts, the decision rule is: if a file is present at the originating site, use the copy there; otherwise, choose uniformly from among the sites that have remote copies of the files. The transaction manager also models the details of the commit and abort protocols.

(iii) The concurrency control manager:
It deals with the implementation of the concurrency control algorithms. In this study, this module is not fully implemented. The effect of this is dependent on algorithm that chooses during designing the system.

(iv) The resource manager:
The resource manager models the physical resources like CPU, Disk, and files etc for writing to or accessing data or messages from them.

(v) The sink:
The sink deals for collection of statistics on the completed transactions.

(vi)The Network Manager:
The network manager encapsulates the model of the communications network. It is assuming a local area network system, where the actual time on the wire for messages is negligible.

## 3. EXECUTION MODEL AND SIMULATION PARAMETERS

The execution model is discussed below. A common model of a distributed transaction is that there is one process, called as Master, which is executed at the site where the transaction is submitted, and a set of processes, called Cohorts, which executes on behalf of the transaction at these various sites that are accessed by the transaction. In other words, each transaction has a master process that runs at its site of origination. The master process in turn sets up a collection of cohort's processes to perform the actual processing involved in running the transaction. When cohort finishes executing its portion of a query, it sends an execution complete message to the master. When the master received such a message from each cohort, it starts its execution process.

When a transaction is initiated, the set of files and data items that, it will access are chosen by the source. The master is then loaded at its originating site and initiates the first phase of the protocol by sending PREPARE (to commit) messages in parallel to all the cohorts. Each cohort that is ready to commit, first force-writes a prepared log record to its local stable storage and then sends a YES vote to the master. At this stage, the cohort has entered a prepared state wherein it cannot unilaterally commit or abort the transaction but has to wait for final decision from the master. On other hand, each cohort that decides to abort force-writes an abort log record and sends a NO vote to the master. Since a NO vote acts like a veto, cohort is permitted unilaterally abort the transaction without waiting for a response from the master.

After the master receives the votes from all the cohorts, it initiates the second phase of the protocol. If all the votes are YES, it moves to a committing state by force-writing a commit log record and sending COMMIT messages to all the cohorts. Each cohort after receiving a COMMIT message moves to the committing state, force-writes a commit log record, and sends an acknowledgement (ACK) message to the master. If the master receives even one NO vote, it moves to the aborting state by force writing an abort log record and sends ABORT messages to those cohorts that are in the prepared state. These cohorts, after receiving the ABORT message, move to aborting state, force-write an abort log record and send an ACK message to the master. Finally, the master, after receiving acknowledgement from all the prepared cohorts, writes an end log record and then forgets and made free the transaction. The statistics are collected in the Sink [Jayant et al'90,92,96]. The database is modeled as a collection of DBsize pages that are uniformly distributed across all the NumSites sites. At each site, transactions arrive under Poisson stream with rate ArrivalRate, and each transaction has an associated firm deadline. The deadline is assigned using the formula

$$DT = AT + SF * RT \qquad (1)$$

Here DT, AT, SF and RT are the deadline, arrival rate, Slack factor and resource time respectively, of transaction T. The Resource time is the total service time at the resources that the transaction requires for its execution. The Slack factor is a constant that provides control over the tightness or slackness of the transaction deadlines.

In this model, each of the transaction in the supplied workload has the structure of the single master and multiple cohorts. The number of sites at which each transaction executes is specifying by the Fileselection time (DistDegree) parameter. At each of the execution sites, the number of pages accessed by the transaction's cohort varies uniformly between 0.5 and 1.5 times CohortSize. These pages are chosen randomly from among the database pages located at that site. A page that is read is updated with probability of WriteProb. Summary of the simulation parameter is given in table I.

## Parameter Settings

The values of the parameter set in the simulation are given in table II. The CPU time to process a page is 10 milliseconds while disk access times are 20 milliseconds.

TABLE1. SIMULATION PARAMETERS

| Parameters | Description |
|---|---|
| NumSites orSelectfile | Number of sites in the Database |
| Dbsize | Number of pages in the database. |
| ArrivalRate | Transaction arrival rate/site |
| Slackfactor | Slack factor in Deadline formula |
| FileSelection Time | Degree of Freedom (DistDegree) |
| WriteProb | Page update probability |
| PageCPU | CPU page processing time |
| PageDisk | Disk page access time |
| TerminalThink | Time between completion of 1 transaction & submission of another |
| Numwrite | Number of Write Transactions |
| NumberReadT | Number of Read Transactions |

TABLE II. VALUES OF SIMULATION MODEL PARAMETERS

| Parameters | Set Values | Parameters | Set Values |
|---|---|---|---|
| NumSites | 8 | FileSelection Time | 3 |
| Dbsize | vary(max.2400) | PageCPU | 10ms |
| ArrivalRate | 6 to 8 job/sec | PageDisk | 20ms |
| Slackfactor | 4 | TerminalThink | 0 to 0.5 sec |
| WriteProb | 0.5 | | |

## 4. EXPERIMENTS AND RESULTS

The experiment has been performed using different simulation language like C++Sim, DeNet etc in reports [13,14,15]. For this study, GPSS World is used as a simulator [18]. Literatures are also collected from several recent studies [19, 1, 20, 21,22, 23]. The study for performance evaluation starts by first developing a base model. Further experiments were constructed around the base model experiments by varying a few parameters and process of execution at a time.

The performance metric of the experiments is MissPercent that is the percentage of input transaction that the system is unable to complete before their deadline. The MissPercent values in range of 0% to 20% are taken to represent system performance under "Normal" loads, while ranges of 21% to 100% represent system performance under "heavy" loads. The study analyzes the performance of the system under different workload with varying the arrival rate of the transaction, dynamic slack factors, execution mode etc. The study analyzed the performance using this new concept of 'intelligent agent' technique. Only the statistically significant results are discussed. The experimental results are discussed below. This following section discusses the statistical results of this simulation under different environments.

### 4.1. Comparison of Centralized and Distributed systems

This experiment compares the performance of the system under centralized and distributed. The distributed systems have higher percentage of miss Transactions than centralized system. This higher miss percentage is due to distance between cohorts. This leads to design of a new perfect distributed commit processing protocol to have a real-time committing performance.

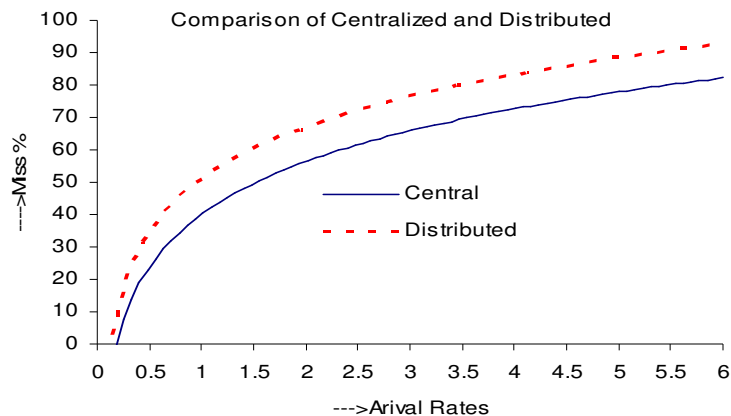

Figure 1 Comparison of Centralize and distributed performances

### 4.2.Impact of distribution methods

This experiment is conducted to know the impact of difference between distribution methods to the performance of the system. As an example, we take Exponential distribution and Poisson distribution. The assignment and committing of transactions to cohorts are passed under scheduler using Exponential distribution and Poisson distribution and the statistics of the simulation outputs are noted. The Exponential give more uniform assignment and committing of transactions than Poisson. Poisson throws higher numbers of transactions giving more collisions of transactions and large number miss percentage of transactions than Exponential. So on many experiments of such similar types can be conducted by using more different distribution rules.

### 4.3. Impact execution mode:Distribution and Parallel

This experiment compares the output of the system putting the cohorts in parallel with that of distribution execution. From this we can conclude following points. Parallel execution of the cohorts reduces the transaction response time. The time required for the commit processing is partially reduced. This is because the queuing time is shorted in parallel and so there are much fewer chances of a cohort aborting during waiting phase. In the following figure, d1, d2 and d3 are representing cohorts connected in distributed and p1, p2 and p3 represents those connected in parallel.

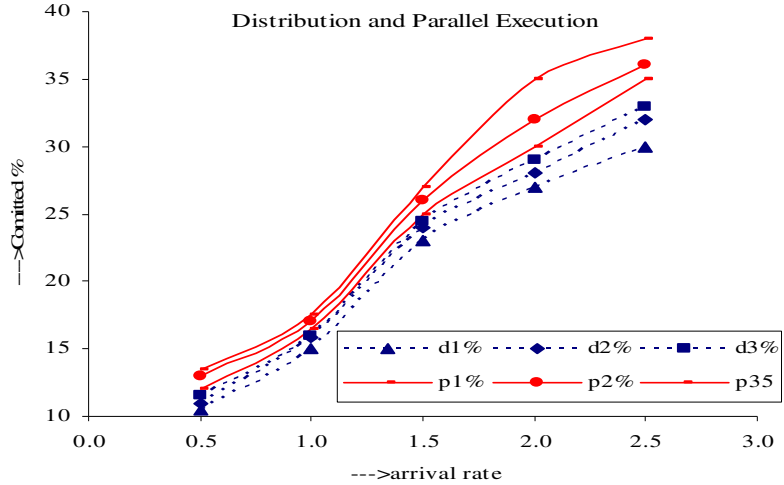

Figure 2 Impact of execution mode-distribution and Parallel

## 4.4. Impact of slack to Throughput

In this set of experiments, the impact of slack factor to observed on the throughput of the system. The throughput initially increases with increase in slack factor –statically or dynamically. But it drops rapidly at very high loads. Here we collected only statically results, not the exact results. Still there are lots more to study required about other parameters to improve the throughput of the overall system. The following figure 3 shows impact of slack factors to throughput of all 8 sites.

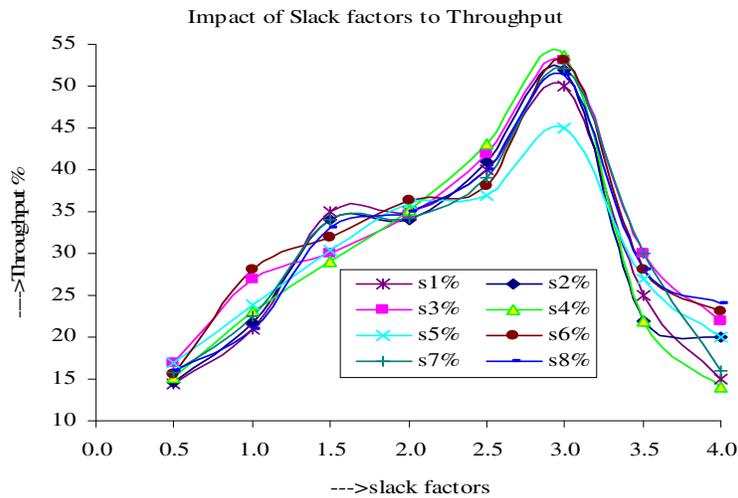

Figure3. Impact of Slack factors to Throughput

### 4.5. Dynamic Management

The transactions can be manage in many differet ways. In most of the earlier works done the transcations are management in static ways, where the parameter values are fixed to a constant value during execution of the experiments. Here in our work, we are introducing experiment done with dynamic manegments of transacions, where the values of the paremeters are changes or adjust automatically depending on the requirements during the execution the experiment. Here we are conducting two different experiments.

### 4.5.1. Impact of Dynamic Intelligent agent

A dynamic intelligent agent is introduced to take sensory input from the environment, and produces as output actions that affect it. The agents act in rapidly changing, unpredictable or open environments and where their action can fail are known to the agents. Main advantages for the use of the mobile agents are the increased the autonomy of the system in many ways like flexibility to topological changes, to load balance changes. After knowing the status of transactions to be fail, extra slack factor can be apply to safe the transaction. The fig 4 shows the performance comparison between normal/base model and impact of the intelligent agent. By adding such agent can improved the performance of the system.

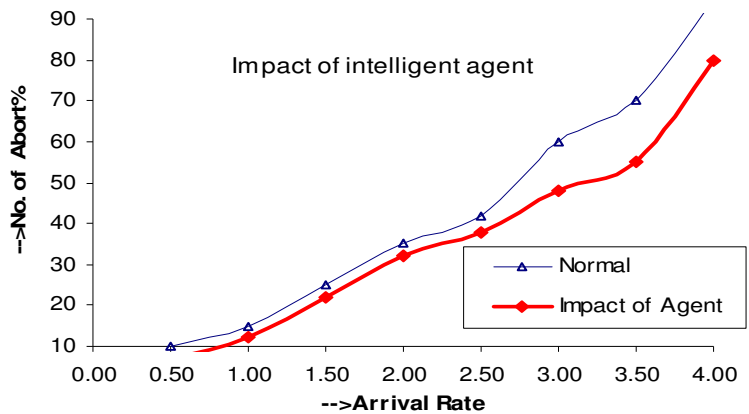

Figure 4. Impact of intelligent agent

### 4.5.2. Impact of dynamic Slack factor

This set of experiment is conducted to safe some of the transactions before killing or aborting them. The Slack factor is provided to control over the tightness or slackness of the transaction deadlines. Normally the system aborts all transactions which are unlikely to be completed before their deadlines. If the slack factor value is nearly negative, it aborts the transaction and removed it from the queue.

After computing the slack value of all transactions, the system will know possible total number of transactions which have +vet and -vet slack values. If there is large number of transaction with +vet slack values means that the system will have some bonus time. If the system is not having firmed slack condition, it can alter some of –vet slack value to next higher level to safe some of the killing transactions. In other words, the bonus time remains from the transactions that have +vet slack value are going to used to safe the transaction with -vet slack value. With the static slack value, it gives normal performance. By increasing the slack dynamically, it gives lower values of missed percentage of transactions. So on exchange of slack time between transactions increased the performance of the system.

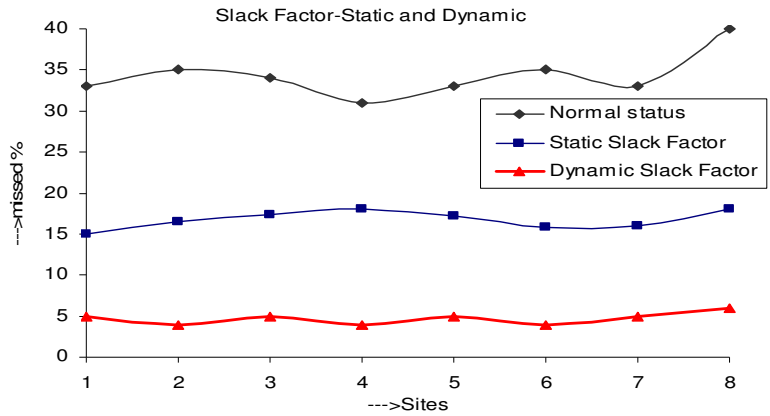

Figure5. Impact of Slack factor to Individual sites

# 5. CONCLUSION

A series of simulation study have been performed to analyze the performance under different transaction management situation such as different workloads, distribution methods, execution mode-Distribution and Parallel, impact of dynamic slack factors to throughput. The scheduling of data accesses are done in order to meet their deadlines and to minimize the number of transactions that missed deadlines.

Parallel execution of the cohorts reduces the transaction response time than that of serial or distributed execution. The time required for the commit processing is partially reduced, because the queuing time is shorted in parallel and so there are much fewer chances of a cohort aborting during waiting phase. The throughput initially increases with increase in slack factor. But it drops rapidly at very high work loads. The slack factors can be providing by static or dynamics ways.

The new approach dynamic management either dynamic intelligent agent or dynamic slack management gives a significant improvement to the performance of the system. Dynamic intelligent agent keeps tracks of timing of the transactions to help them from aborts. This agent gives advance information about the remaining execution time of the transactions. This helps the system to inject extra time to such transactions. In all the conditions the arrival rate of transaction plays a major role in reducing number of miss percentage and improved performance.

# REFERENCE

[1] Silberschatz, Korth, Sudarshan,2002, *Database system concept*,4th (I.E), McGraw-Hill Pub. 698-709,903

[2] Gray. J,1978,"Notes on Database Operating Systems", Operating Systems:An Advanced Course, *Lecture notes in Computer Science*

[3] Mohan, C, Lindsay B and Obermark 1986, Transaction Management in the R*  Distributed Database Management Systems, *ACM TODS*, 11(4).

[4] Lampson B and Lomet D,1993, "A new Presumes Commit Optimization for Two phase Commit", *Pro.of 19th VLDB Conference.*

[5] Oszu M, Valduriez P,1991, *Principles of Distributed Database Systems*, Prentice-Hall.

[6] Kohler W, 1981,A survey of Techniques for Synchronization and Recovery in Decentralized Computer System, *ACM Computing Surveys*, 13(2)

[7] Nystrom D, Nolin M,2006, Pessimistic Concurrency Control and Versioning to Support Database Pointers in Real-Time Databases, *Proc. 16th Euromicro Conf. on Real-Time Systems*

[8] Ramamritham,Son S. H, and DiPippo L,2004, Real-Time Databases and Data Services, *Real-Time Systems J.*, vol. 28, 179-216.

[9] Robert A and Garcia-Molina H,1992, Scheduling Real-Time Transactions, *ACM Trans. on Database Systems*, 17(3).

[10] Levy E., Korth H and Silberschatz,1991,An optimistic commit protocol for distributed transaction management, *Pro.of ACM SIGMOD Conf.*

[11] Jayant. H, Carey M, Livney,1992, "Data Access Scheduling in Firm Real time Database Systems", *Real Time systems Journal*, 4(3)

[12] Jayanta Singh and S.C Mehrotra, 2006,"Performance analysis of a Real Time Distributed Database System through simulation" *15th IASTED International Conf. on APPLIED SIMULATION & MODELLING, Greece*

[13] Jayanta Singh and S.C Mehrotra,2009 "A study on transaction scheduling in a real-time distributed system",*EUROSIS's Annual Industrial Simulation Conference, UK.*

[14] Jayant H.  1991, "Transaction Scheduling in Firm Real-Time Database Systems", *Ph.D. Thesis, Computer Science Dept. Univ. of Wisconsin, Madison.*

[15] Jayant H. Carey M and Livney M,1990 "Dynamic Real-Time Optimistic Concurrency Control", *Proc. of 11th IEEE Real-Time Systems Symp.*

[16] Jayant H., Ramesh G. Kriti.R, S. Seshadri, "Commit processing in Distributed Real-Time Database Systems", Tech. Report-TR-96-01, Pro. Pro. Of 17th  *IEEE Real-Time Systems Symposium, USA,1996*

[17] Han Q, 2003, Addressing timeliness /accuracy/ cost tradeoffs in information collection for dynamic environments, *IEEE Real-Time System Symposium,Cancun, Mexico*

[18] Minutesmansoftware, *GPSS world*, North Carolina,  U. S. A. 2010.

[19] Xiong M. and Ramamritham K.,2004, Deriving Deadlines and Periods for Real-Time Update Transactions, *IEEE Trans. on Computers*, vol. 53,(5).

[20] Gustavsson S and Andler S 2005, Decentralized and continuous consistency management in distributed real-time databases with multiple writers of replicated data, *Workshop on parallel and distributed real-time systems, Denver, CO*

[21] Xiong M, Han S., and Lam K,2005,  A Deferrable Scheduling for Real-Time Transactions Maintaining Data Freshness, *IEEE Real-Time Systems Symposium.*

[22] Jan Lindstrom,2006 "Relaxed Correctness for Firm Real-Time Databases," rtcsa,pp.82-86, *12th IEEE International Conference on Embedded and Real-Time Computing Systems and Applications (RTCSA'06).*

[23] Idoudi, N.  Duvallet, C.  Sadeg, B.  Bouaziz, R.  Gargouri, F,2008, Structural Model of Real-Time Databases: An Illustration, *11th IEEE International Symposium on Object-Oriented Real-Time Distributed Computing (ISORC 2008).*

# BIOGRAPHY

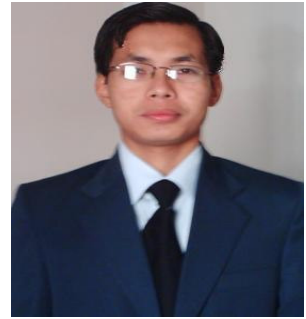

Y. JAYANTA SINGH is working as a Lecturer in Faculty of Information technology, 7[th] October University, Misurata, (Libya). He obtained his Ph.D in Computer Science from Dr. B.A.Marathwada University, (India) in 2004. He had worked with Keane (Canada), Skyline University College (UAE). His areas of interest are Distributed Real time, Software Engineering and Simulation and modeling etc.

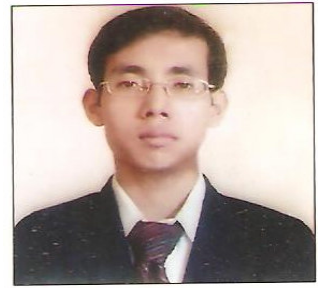

YUMNAM SOMANANDA SINGH is working in Faculty of Computer Application, Institute of Management Studies & I. T(IMSIT) Aurangabad, India. He has Master and M.Phil in Computer Science. He is working as a Ph.D scholar. His areas of interest are Data processing, Distributed Database, Image processing etc.

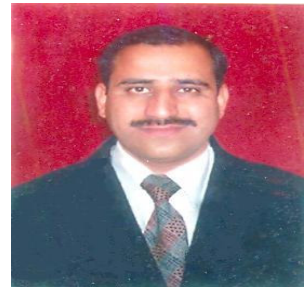

ASHOK GAIKWAD is working as an Associate Professor and Head of Department in Faculty of Computer Application, Institute of Management Studies & I. T (IMSIT) Aurangabad, India. He obtained his Ph.D in Computer Science from Dr. B.A.Marathwada University, India, 2007. His areas of interest are distributed Real time, Software Engineering and Digital Image processing etc. (drashokgaikwad@gmail.com).

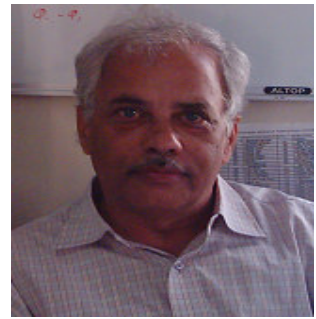

S. C MEHROTRA, F.N.A,Sc., FIETE  is working as a professor in Dr. Babasaheb Ambedkar Marathwada University, Aurangabad (India). He received his master degree in Physics from Allahabad University (India) in 1970 and Ph.D. in Physics from Austin (USA) in 1975. He is recipient of Welch Foundation Fellowship (1975), Alexander Von Humboldt Foundation Fellowship (1983-85), FOM (Netherland). He has published more than 150 papers in areas of Time Domain Reflectometery, Speech Processing, and Network Simulation etc